\documentclass[a4paper,reqno]{amsart}
\usepackage[text={5.2in,8in},centering]{geometry}

\usepackage[usenames,dvipsnames,svgnames,table]{xcolor}
\definecolor{citegreen}{rgb}{0.00,0.70,0.30}
\usepackage[colorlinks=true,
            linkcolor=blue,
            urlcolor=blue,
            citecolor=blue]{hyperref}

\usepackage{amsrefs}
\usepackage{amssymb}

\usepackage{amsmath}
\usepackage{amsthm}

\usepackage{mathrsfs}

\usepackage{appendix}

\usepackage{xspace}

\usepackage{enumerate}
\usepackage{graphicx}

\usepackage{pgf}
\usepackage{tikz}
\usetikzlibrary{patterns}

\numberwithin{equation}{section}
\theoremstyle{plain}
\newtheorem{theorem}{Theorem}
\newtheorem{lemma}{Lemma}[section]

\newtheorem{prop}[lemma]{Proposition}
\newtheorem{cor}[lemma]{Corollary}

\theoremstyle{remark}
\newtheorem{remark}[theorem]{Remark}

\newtheorem*{quest*}{Question}
\newtheorem*{remark*}{Remark}
\theoremstyle{definition}
\newtheorem{definition}{Definition}[section]

\newtheorem*{notation*}{Notation}
\newtheorem*{notations*}{Notations}
\providecommand{\B}{\mathbf}

\providecommand{\D}{\mathbb}

\newcommand{\ee}{\mathrm{e}}

\providecommand{\esm}[1]{\D{E}\left[ #1 \right]}

\DeclareMathOperator{\dist}{dist}
\DeclareMathOperator{\diam}{diam}
\DeclareMathOperator{\supp}{supp}
\DeclareMathOperator{\card}{card}

\DeclareMathOperator{\one}{\mathbf{1}}

\DeclareMathOperator{\out}{out}

\definecolor{redd}{rgb}{0.95,0.2,0.2}
\definecolor{gris}{rgb}{0.9,0.9,0.9}
\definecolor{greenn}{rgb}{0.1,0.6,0.2}

\definecolor{cmgray}{rgb}{0.7,0.7,0.7}

\definecolor{cmblue}{rgb}{0.2,0.5,0.8}

\DeclareMathAlphabet{\mathpzc}{OT1}{pzc}{m}{it}
\def\cond{\,\big|\,}

\def\be{\begin{equation}}
\def\ee{\end{equation}}
\def\ba{\begin{array}{l}}
\def\ea{\end{array}}
\def\bal{\begin{aligned}}
\def\eal{\end{aligned}}

\def\ble{\begin{lemma}}
\def\ele{\end{lemma}}

\def\bthm{\begin{theorem}}
\def\ethm{\end{theorem}}

\def\bco{\begin{cor}}
\def\eco{\end{cor}}

\def\bpr{\begin{prop}}
\def\epr{\end{prop}}

\def\bre{\begin{remark}}
\def\ere{\end{remark}}

\def\btm{\begin{theorem}}
\def\etm{\end{theorem}}

\def\bde{\begin{definition}}
\def\ede{\end{definition}}

\def\eu{{\mathrm{e}}}

\def\ffi{\varphi}

\def\BLam{\boldsymbol{\Lambda}}

\def\BDelta{\boldsymbol{\Delta}}

\def\bcubeout{\boldsymbol{\Lambda}^{\text{out}}}

\def\half{\frac{1}{2}}
\def\shalf{{\textstyle{\frac{1}{2}}}}
\def\third{\frac{1}{3}}

\def\quart{\frac{1}{4}}
\def\squart{{\textstyle{\frac{1}{4}}}}

\def\EmNS{$(E,m)$-{\rm{NS}}\xspace}
\def\EmS{$(E,m)$-{\rm S}\xspace}

\def\eps{\epsilon}
\def\lam{\lambda}

\def\pt{\partial}
\def\Const{\mathrm{Const\,}}

\def\pr#1{\mathbb{P}\left\{ #1 \right\}}
\def\esm#1{\D{E}\left[\, #1\, \right]}

\newcommand{\vertii}[1]{{\big\vert\kern-0.25ex\big\vert #1
    \big\vert\kern-0.25ex\big\vert\kern-0.25ex}}

\newcommand{\vertiii}[1]{{\big\vert\kern-0.25ex\big\vert\kern-0.25ex\big\vert #1
    \big\vert\kern-0.25ex\big\vert\kern-0.25ex\big\vert}}

\newcommand{\dnorm}[1]{{\big\| #1 \big\|^{\curlywedge}}}

\def\CbetaNR{{$(E,\beta)${\rm-CNR }}}

\def\diy{\displaystyle}

\def\BC{\mathbf{C}}
\def\bcell{\mathbf{C}}
\def\Bzero{\mathbf{0}}

\def\bcubkoneu{\bcube_{L_{k+1}}(\Bu)}

\def\Cgeom{C^{\rm GRI}}

\def\lam{\lambda}
\def\om{\omega}
\def\th{\theta}
\def\ffi{\varphi}

\def\Om{\Omega}
\def\tOm{\widetilde{\Omega}}

\def\Sigmani{\Sigma^{\rm ni}}

\def\vempty{\varnothing}

\def\EmS{{\rm$(E,m_N)$-S}\xspace}
\def\EmNS{$(E,m_N)$-NS\xspace}

\def\ER{{\rm$(E,\beta)$-R}\xspace}
\def\ENR{$(E,\beta)$-NR\xspace}

\def\Egood{{\rm$(E,m_N,K)$-good}\xspace}
\def\Ebad{$(E,m_N,K)$-bad\xspace}

\def\SRCM{\textsf{(SRCM)}\xspace}

\def\Vone{\textsf{(V)}\xspace}

\def\Uone{\textsf{(U)}\xspace}

\def\bcube{{\boldsymbol{\Lambda}}}
\def\bcubeN{\boldsymbol{\Lambda}^{N}}

\def\tbcube{\widetilde{\boldsymbol{\Lambda}}}

\def\bcL{\boldsymbol{\mathcal{L}}}

\def\fS{\mathfrak{S}}
\def\fF{\mathfrak{F}}

\def\BF{\mathbf{F}}

\def\BG{\mathbf{G}}
\def\BGni{\mathbf{G}^{\mathrm{ni}}}

\def\BH{\mathbf{H}}
\def\BHni{\mathbf{H}^{\mathrm{ni}}}

\def\BP{\mathbf{P}}
\def\BU{\mathbf{U}}
\def\BV{\mathbf{V}}

\def\ibA{\boldsymbol{A}}
\def\ibB{\boldsymbol{B}}

\def\BPsi{\boldsymbol{\Psi}}

\def\bcX{\boldsymbol{\mathcal{X}}}

\def\bcZ{\boldsymbol{\mathcal{Z}}}

\def\bcZN{\boldsymbol{\mathcal{Z}}^{N}}

\def\SS#1{{\textsf{S}$\left(#1\right)$}}

\def\bcubeLkone{\BLam_{L_{k+1}}}

\def\csE{\mathscr{E}}

\def\csS{\mathscr{S}}

\def\cB{\mathcal{B}}

\def\cJ{\mathcal{J}}

\def\cS{\mathcal{S}}

\def\cT{\mathcal{T}}
\def\cV{\mathcal{V}}
\def\cZ{\mathcal{Z}}

\def\bball{\mathbf{B}}
\def\bballN{\mathbf{B}^{(N)}}

\def\Bx{\mathbf{x}}
\def\hBx{\widehat{\mathbf{x}}}

\def\By{\mathbf{y}}
\def\Bz{\mathbf{z}}
\def\Bu{\mathbf{u}}
\def\Bv{\mathbf{v}}

\def\rc{\mathrm{c}}
\def\rd{\mathrm{d}}

\def\rr{\mathrm{r}}

\def\rC{\mathrm{C}}

\def\rM{\mathrm{M}}
\def\rP{\mathrm{P}}
\def\rQ{\mathrm{Q}}
\def\rS{\mathrm{S}}

\def\tm{{\widetilde{m}}}

\def\DA{\mathbb{A}}

\def\DP{\mathbb{P}}
\def\DR{\mathbb{R}}
\def\DZ{\mathbb{Z}}
\def\DN{\mathbb{N}}

\def\tto#1{\smash{\mathop{\,\,\,\, \longrightarrow \,\,\,\, }\limits_{#1}}}

\begin{document}
\title[Localization in a continuous multi-particle Anderson model]
{Exponential decay of eigenfunctions\\in a continuous multi-particle Anderson model\\with sub-exponentially decaying interaction}
\author{Victor Chulaevsky}
\date{}
\begin{abstract}
This short note is a complement to our recent paper [2] where we established strong dynamical localization
for a class of multi-particle Anderson models in a Euclidean space
with an alloy-type random potential and a sub-exponentially decaying interaction of infinite range.
We show that the localized eigenfunctions at low energies actually decay exponentially fast.
This improves the results by Fauser and Warzel who established sub-exponential decay of eigenfunctions
in presence of a sub-exponentially decaying interaction.
\end{abstract}
\maketitle
\section{Introduction} \label{sec:intro}
\subsection{The model and the main goal}

We study a multi-particle Anderson model in $\DR^d$ with long-range interaction
and subject to an external random potential of the so-called alloy type.
The Hamiltonian $\BH(\om)= \BH^{(N)}(\omega)$ is a random Schr\"{o}dinger operator of the form
\begin{equation}\label{eq:def.H}
\BH(\om) = -\frac{1}{2}\BDelta + \BU(\Bx)+\BV(\omega;\Bx)
\end{equation}
acting in $L^2\big( (\DR^{d})^N\big)$. To stress the dependence on the number of particles, $n\ge 1$, omitting a less
important parameter $d $ ($=$ the dimension of the $1$-particle configuration space), we denote
$$
\bcX^N := \big(\DR^d\big)^N,  \;\; \bcZ^N := \big(\DZ^d\big)^N \hookrightarrow \big(\DR^d\big)^N, \;\; N\ge 1.
$$
The points $\Bx=(x_1,\ldots,x_N)\in \bcX^N$ represent the positions of the $N$ quantum particles evolving simultaneously
in the physical space $\DR^d$.
In \eqref{eq:def.H},
$\BDelta$ stands for the Laplacian in $(\DR^d)^N$ (or, equivalently, in $\DR^{Nd}$). The interaction energy
operator $\BU$ acts as multiplication by a function $\Bx \mapsto \BU(\Bx)$. Finally, the potential energy $\BV(\omega;\Bx)$
(unrelated to the inter-particle interaction)
is the operator of multiplication by a function
\begin{equation}\label{eq:def.V}
\Bx\mapsto V(x_1;\omega)+ \cdots + V(x_N;\omega),
\end{equation}
where $x\in\DR^d\mapsto V(x;\omega)$ is a random external field potential assumed to be of the form
\begin{equation}\label{eq:defalloy}
V(x;\omega)=\sum_{a\in\D{Z}^d}\cV_a(\omega)\, \ffi(x-a).
\end{equation}
Here and below $\cV_a$, $a\in\D{Z}^d \hookrightarrow \DR^d$, are IID (independent and identically
distributed) real random variables on some probability space $(\Omega,\fF,\DP)$ and
$\ffi:\DR^d\to\DR$ is usually referred to as a scatterer (or ``bump'') function.

More precise assumptions will be specified below.

For the motivation and bibliographical remarks, see \cite{C14b}, where it is shown that the random Hamiltonian
$\BH(\om)$, satisfying the hypotheses \Vone and \Uone formulated below, features strong dynamical localization
with eigenfunction correlators decaying at sub-exponential rate ($r \mapsto \eu^{-\nu r^\kappa}$, with some
$\nu,\kappa>0$).
\vskip1mm

By the RAGE theorems, it follows from the results of \cite{C14b} that with probability one, $\BH(\om)$ has pure point spectrum
in an energy zone near the bottom of the spectrum, and all respective eigenfunctions decay at a sub-exponential rate (perhaps,  faster).
In the present paper, we complement these results by proving that in the localization zone, the eigenfunctions actually
decay exponentially fast at infinity, even if the interaction potential $U(r)$ decays at sub-exponential rate $\eu^{-r^\zeta}$,
with arbitrarily small $\zeta>0$.
For the model considered here, this improves the results by Fauser and Warzel \cite{FW14} who proved (prior to \cite{C14b})
sub-exponential decay of eigenfunction correlators (hence, sub-exponential decay of EFs),
in a more general class of $N$-particle Hamiltonians with alloy potentials and a sub-exponentially decaying interaction.

We work only with a particular class of alloy potentials, called in \cite{C14b} \emph{flat tiling} alloys, for two reasons.
\begin{itemize}
  \item This paper completes the results of \cite{C14b}, where we focused on strong dynamical localization.

  \item The main strategy of proof of exponential localization used here relies essentially on the fixed-energy Multi-Scale Analysis (MSA),
  and a relative simple, "soft" derivation of the energy-interval bounds, required for spectral localization with exponentially decaying eigenfunctions,
  is based on a special kind of eigenvalue concentration (EVC) bound which has been proved so far only for a limited class of
  multi-particle Anderson Hamiltonians.
\end{itemize}

On the other hand, the extension of the techniques of Ref.~\cite{CBS11} (energy-interval multi-particle MSA) to the continuous
Anderson models with sub-exponentially decaying interaction can be carried out for a much larger class of alloy potentials
and under much weaker assumptions on the regularity of the probability distribution of the scatterers' amplitudes.
Specifically, it suffices to assume log-H\"{o}lder continuity of the marginal probability distribution function; such models are known
to be beyond the reach of the Fractional Moment Method, even in the one-particle localization theory.
Moreover, the EVC bound proved recently by Klein and Nguyen \cite{KN13b} and based on the Unique Continuation Principle
for spectral projections (UCPSP; cf. \cite{Kle13})
applies to alloys even without the complete covering condition (assumed in \cite{FW14}), let alone the flat tiling condition
(assumed both in \cite{C14b} and in the present  paper)
which is, in a manner of speaking, a "perfect complete covering" condition.
However, the energy-interval MPMSA is more complex than its fixed-energy counterpart, and for this reason it cannot
be included into the present manuscript. Compared to Ref. \cite{CBS11}, the main modification required in the case of
an infinite-range interaction is encapsulated in an analog of Lemma \ref{lem:WITRONS.subexp}.

Recall that Fauser and Warzel \cite{FW14} proved exponential decay of eigenfunctions (with low-lying eigenvalues) in a multi-particle
continuous  alloy model with \emph{exponentially} decaying interaction, under the complete covering condition.


\subsection{Basic geometric objects and  notations}

\begin{center}
\begin{figure}
\begin{tikzpicture}
\begin{scope}[scale=0.40]
\clip (-7.5,-6.0) rectangle ++(15.0,11.5);

\fill[color=black!80!white!50, line width = 0.5, dotted] (-0.47, -0.47) rectangle ++(0.94, 0.94);

\foreach \i in {-5.5, -4.5, ..., 4.5}
{
\fill[color=black!50!white!50] (\i+0.03, -5.47) rectangle ++(0.94, 0.94);
\fill (\i+0.5+0.03, -5.47+0.5) circle  (0.1);
\fill[color=black!50!white!50] (\i+0.03, 4.53) rectangle ++(0.94, 0.94);
\fill (\i+0.5+0.03, 4.5+0.5) circle  (0.1);

\foreach \j in {-5.0, -4.0,..., 5.0} \fill (\i+0.5, \j) circle  (0.045);

}

\foreach \i in {-4.5, -3.5, ..., 3.5}
{
\fill[color=black!50!white!50] (-5.47, \i) rectangle ++(0.94, 0.94);
\fill[color=black!50!white!50] (4.53, \i) rectangle ++(0.94, 0.94);

\fill (-5.0, \i+0.5) circle  (0.1);
\fill ( 5.0, \i+0.5) circle  (0.1);

}

\end{scope}
\end{tikzpicture}
\caption{\emph{Cubes, cells and boundaries}.
Here $d=1$, $N=2$, $L=5$, $\Bu=\Bzero = (0,0)$, so the cube $\bcube_L^N(\Bu)$
is the square of diameter $2L+1=11$ in the max-norm, covered by $(2L+1)^2$
squares of diameter $1$ centered at integer points in $\bcZ^2 \cong \DZ^2$.
The $4\cdot 2L$ light gray squares cover the boundary belt $\bcube_L^{\out}(\Bzero)=\{\By\in\DR^2:\, |\By|\in(L-\half, L+\half)\}$.
The larger dots represent the boundary $\pt^- \bball_L(\Bzero)$ of the lattice cube
$\bball_L(\Bzero) = \bcube_L(\Bzero) \cap \DZ^2$, with  $|\bball_L(\Bzero)| = (2L+1)^2$
and $\diam \bball_L(\Bzero) = 2L$.
}
\end{figure}
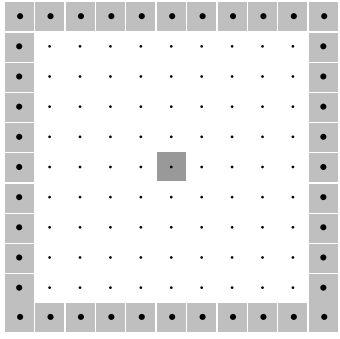
\end{center}

We fix from the beginning an integer $N^* \geq 2$, which can be arbitrary large, and work in Euclidean spaces of the
form $(\DR^{d})^N\cong\DR^{Nd}$, $1\le N \le N^*$. A configuration of $N\ge 1$ distinguishable particles
in $\DR^d$ is represented by (and in our paper, identified with) a vector $\Bx=(x_1, \ldots, x_N)\in(\DR^d)^N$,
where $x_j$ is the position of the $j$-th particle.
In general, boldface notations are reserved for "multi-particle" objects (Hamiltonians, resolvents, cubes, etc.).

Since we work with the alloy-type external random potentials, an important role is played by the integer lattices
$\DZ^d \hookrightarrow \DR^d$ and $(\DZ^d)^N \hookrightarrow (\DR^d)^N$.

All Euclidean spaces will be endowed with the max-norm denoted by $|\,\cdot\,|$, so that $|\Bx|=\max_i |x_i|$. We will consider
$Nd$-dimensional cubes of integer edge length in
$(\DZ^d)^N$ centered at lattice points $\Bu\in(\DZ^d)^N \hookrightarrow (\DR^d)^N$ and with edges parallel to the co-ordinate axes.
The open cube of edge length $2L+1$ centered at $\Bu$ is denoted by $\bcube_L(\Bu)$; in the max-norm
it represents the open ball of radius $L+\half$ centered at $\Bu$:
\begin{equation}\label{eq:BLam}
\bcube_L(\B{u}) =\{\Bx\in\bcX^N: \;|\Bx - \Bu| < L + \shalf\}.
\end{equation}
The lattice counterpart for $\bcube_L(\Bu)$ is denoted by $\bball_L(\Bu)$:
$$
\bball_L(\Bu) = \bcube_L(\Bu) \cap \bcZ^N = \{ \Bx\in\bcZ^N:\, |\Bx - \By| \le L\}
; \quad \Bu\in \bcZ^N.
$$
$\bball_L(\Bu) $ is the lattice ball of radius $L$ (not $L+\half$, as in the case of $\bcube_L(\Bu$)),
so by a slight abuse of terminology, we often refer to $L$ as the radius of both $\bball_L(\Bu)$
and $\bcube_L(\Bu)$.

We call a \emph{cell} a closed cube of diameter $1$ centered at a lattice point $\Bu\in\bcZN$:
$$
\BC(\Bu) = \{\By\in\bcX^N: \; |\By - \Bu| \le \shalf \} .
$$
The union of all cells $\BC(\Bu)$, $\Bu\in\bcZN$, covers the entire Euclidean space $\bcX^N$.
Moreover, for any $\Bu\in\bcZ^N$, denoting by $\overline{\ibA}$ the closure of the set $\ibA\subset\bcX^N$, we have
$$
\overline{\bcube_L(\Bu)} = \bigcup_{ \By \in \bball_L(\Bu)} \bcell(\By).
$$

The diameters appearing in our formulae are relative to the max-norm;
the cardinality of various sets $A$ (usually finite) will be often denoted by $|A|$. We have
$$
\diam \bcube_L(\Bu) = 2L+1, \diam \bball_L(\Bu) = 2L, \;\; | \bball_L(\Bu)| = (2L+1)^{Nd} \le (3L)^{Nd}.
$$

The indicator function of a set $A$ is denoted in general by $\one_A$, but for the indicators of the cells
we use a shorter notation, $\chi_\Bx := \one_{\bcell(\Bx)}$.

The standard application of the Geometric Resolvent Inequality for the continuous Schr\"{o}dinger operators
(cf. Sect.~\ref{ssec:dominated.decay}) involves the boundary belt of a cube $\bcube_L(\Bu)$:
\be
\bcube^{\out}_L(\Bu) := \bcube_L(\Bu) \setminus \overline{\bcube_{L-1}(\Bu)}
= \left\{\By: |\By - \Bx| \in \left(L-\shalf, L+\shalf \right) \right\}.
\ee
We also define the boundary $\pt^- \bball_L(\Bu)  = \{ \By\in\bball_L(\Bu):\, |\Bu - \By|= L\}$ of the lattice counterpart of
$\bcube_L(\Bu)$ in such a way that (cf. Fig.~1)
\be\label{eq:def.pt.bball}
\bcube^{\out}_{L}(\Bu) \subset \bigcup_{\Bx\in\pt \bball_L(\Bu)} \BC(\Bx).
\ee

\subsection{Symmetrized norm-distance}

The symmetrized norm-distance is much more
natural than the conventional norm-distance in the multi-particle configuration space,
even in a situation where, as in the present paper, the particles are considered distinguishable.
The formal definition is as follows:
$$
\rd^{(N)}_S( \Bx, \By) := \min_{\pi\in\fS_N} | \pi(\Bx) - \By |,
$$
where the elements of the symmetric group $\pi\in\fS_N$ act on $\Bx = (x_1, \ldots, x_N)$ by permutations of the coordinates $x_j$.

\subsection{Interaction potential}

We assume the following:

\Uone $\BU$ is generated by a $2$-body potential $U:\DR_+\to \DR_+$, viz.
$$
\BU(\Bx) = \sum_{1\leq i < j\le N  }U(|x_i - x_j|),
$$
where
\be
0 \le U(r) \le C_U \eu^{-r^\zeta} ,
\ee
for some  $\zeta>0$, $C_U\in(0,\infty)$. We stress that $\zeta>0$ can be \textbf{arbitrarily small}.

\subsection{External random potential}
\label{ssec:cond.on.V}

\begin{center}
\begin{figure}
\begin{tikzpicture}
\begin{scope}[scale=0.40]
\clip (-7.5,-7.0) rectangle ++(17.0,9.0);

\foreach \i in {-7.5, -6.5, ..., -1.0}
{

\foreach \j in {-7.0, -6.0,..., -1.0}
{
\fill[color=gray] (\i+0.05,\j) -- (\i+1-0.05,\j) -- (\i+2-0.05, \j+1-0.05) -- (\i+1+0.05, \j+1-0.05) -- cycle;
\fill (\i+0.99, \j+0.5) circle  (0.1);
}
}

\foreach \i in {1.5, 2.5, ..., 8.0}
{
\foreach \j in {-7.0, -6.0,..., -1.0}
{
\fill[color=gray] (\i+0.05,\j) -- (\i+1-0.05,\j) -- (\i+1-0.05, \j+1-0.05) -- (\i+0.05, \j+1-0.05) -- cycle;
\fill (\i+0.5, \j+0.5) circle  (0.1);
}

}

\end{scope}
\end{tikzpicture}
\caption{\emph{Two examples of flat tiling in $\DR^2$. See Remark \ref{rem:flat.tiling}}.
}
\end{figure}
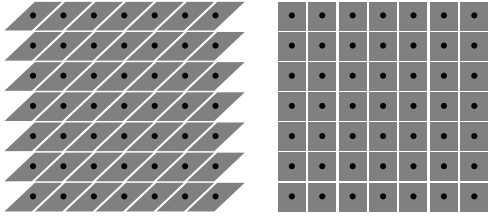
\end{center}

We assume the following conditions to be fulfilled.

\Vone
The external random potential is of alloy type,
\be\label{eq:alloy}
V(x;\om) = \sum_{a\in\cZ} \cV(a;\om) \ffi_a(x - a),
\ee
where $\cV:\cZ\times\Om\to\DR$ is an IID random field on the lattice $\cZ = \DZ^d \hookrightarrow \DR^d$, relative to some
probability space $(\Om,\fF,\DP)$; the expectation will be denoted by $\esm{\cdot}$.

The scatterer (a.k.a. bump) functions $\ffi_a$ have the following property which we call \emph{\textbf{flat tiling}}:
$\diam \supp\, \ffi_a \le \rr_1 < \infty$ and for some $\rC_\ffi \in(0,+\infty)$,
\be\label{eq:covering.condition}
\sum_{a\in\cZ} \ffi_a =  \rC_\ffi \one ,
\ee
up to a subset of zero Lebesgue measure in $\DR^d$.

The random scatterers amplitudes $\cV(\cdot;\om)$ are IID (independent and identically distributed) and
admit a common probability density $p_V$, which is compactly supported, with
$\supp\, p_V$ $ = [0, c_V]$, $c_V>0$, and $p_V$ is strictly positive, bounded and has bounded
logarithmic derivative in the open interval $(0,c_V)$:
\be\label{eq:cond.pV}
\forall\, t\in (0,c_V)\qquad
\begin{cases} 0 < \;p_* \le p_V(t) \le p^* < +\infty, \\ 0 \le p'_V(t)/p_V(t) \le C^* < +\infty .
\end{cases}
\ee

\bre\label{rem:flat.tiling}
For brevity, we assume $\rC_\ffi=1$; this is inessential for the validity of the main results.
Note that the scatterers' supports can overlap: for example, the indicators of the cubes
$\{y\in\DR^d:\, |y-x|\le n/2\}$ of any integer diameter $n\in\DN^*$ form an $n$-fold flat tiling.
\ere

\subsection{The almost sure spectrum}

The exact location of the a.s. spectrum of the $N$-particle Hamiltonian $\BH^N(\om)$ in the entire Euclidean space $(\DR^d)^N$
can be easily found with the help of the classical Weyl criterion.
A minor
modification\footnote{As was pointed out in \cite{C14b}, we cannot apply directly the results of Ref.~\cite{KN13b}
which addresses the models with a finite-range interaction, but in the proof of Proposition \ref{prop:spec.R+},
this is a pure formality.}
of the argument used in the proof of \cite{KN13b}*{Proposition A.1} leads to the following result.

\bpr
\label{prop:spec.R+}
Under the assumptions \Vone and \Uone, $\Sigma(\BH^N_{\bcX^N}(\om)) = [0,+\infty)$ with probability $1$.
\epr

\subsection {Main result}

\begin{theorem}\label{thm:main}
Assume the conditions {\rm\Vone} and {\rm\Uone} and fix an integer $N^* \ge 2$.
Consider the random Hamiltonian
$\BH^N(\om) = - \BDelta + g\BV(\om) + \BU(\Bx)$ of the form \eqref{eq:def.H}, with $g>0$.
There exist $E^*_g>0$, $m^*_g>0$ such that
for all $N\in[1,N^*]$,
$\BH^N(\om)$ has pure point spectrum in $I^*_g = [0, E^*_g]$, and all
its eigenfunctions $\BPsi_j$ with eigenvalues in $I^*_g$ decay exponentially fast at infinity, with rate $\ge m^*_g$.
Specifically, for $\DP$-a.e. $\om$,
every eigenfunction $\BPsi_j(\om)$ (with eigenvalue in $I^*_g$)
and some $r_j(\om)\in(0,+\infty)$, for all $\Bx\in\bcZ^N$ with $|\Bx| \ge r_j(\om)$,
\be\label{eq:thm.Main}
\| \chi_x \BPsi_j \| \le \eu^{ - m^*_g |\Bx| }.
\ee
Furthermore, $E^*_g, m^*_g \to+\infty$ as $g\to +\infty$.
\end{theorem}

We focus essentially on the general case where $g>0$ is not assumed to be large, and omit
$g$ from the formulae; the main mechanism
responsible for the onset of (multi-particle) localization is the Lifshitz tails phenomenon, like in the
one-particle localization theory.
The last assertion of the theorem will follow from Remark \ref{rem:ILS} on the initial length scale estimate.

\section{EVC bounds}

\subsection{Strong Regularity of the Conditional Mean (SRCM)}

The key property of the probability distribution of the random scatterers in the flat tiling model,
resulting in efficient EVC bounds and ultimately, in norm-bounds on the decay of EFCs
established in \cite{C14b}, can be formulated
for a random field $\cV:\cZ\times\Om\to\DR$ on a countable set $\cZ$ and relative to some probability space
$(\Om,\fF,\DP)$. Formally speaking, it does not presume independence or any explicit decay of correlations
of the random field in question.

Introduce the following notation. Given a finite set $Q\subset\cZ$, we set
$\xi_Q(\om) := |Q|^{-1}\sum_{x\in Q} V(x;\om)$ (the sample, or empirical, mean) and $\eta_x(\om) = V(x;\om) - \xi_Q(\om)$
for $x\in Q$ (the "fluctuations" relative to the sample mean). Denote by $\fF_Q$ the sigma-algebra generated by
all the fluctuations $\eta_x$, $x\in Q$.

\vskip1mm

\vskip1mm
\noindent
\SRCM:
\emph{ Given a random field $\cV:\cZ\times\Om\to\DR$ on a countable set $\cZ$,
there exist $C, A \in(0,+\infty)$ such that
for any finite subset $Q\subset\cZ$,
any $\fF_Q$-measurable random variable $\mu$ and all $s\in(0, 1]$, the following
bound holds:
\be\label{eq:CMxi}
\pr{ \xi_Q(\om) \in [\mu(\om), \mu(\om)+s] \cond \fF_Q } \le C\, |Q|^{A} s.
\ee
} 

\bpr[Cf. \cite{C14c}]
\label{prop:SRCM}
Under the assumption {\rm\Vone}, the IID random field $\cV: \cZ\times \Om \to\DR$ satisfies the property {\rm\SRCM} with $A = 2$.
\epr

In Ref.~\cite{C14b}, we used a weaker form of \eqref{eq:CMxi}, called there Regularity of Conditional Mean (RCM).
The main improvement concerns the exponent of $s$ in the RHS: in \cite{C14b} we assumed an upper bound by $C |Q|^{A} s^\th$
for some $\th>0$. The optimal regularity exponent $\th=1$
is required for the proof of Theorem \ref{thm:FEMSA.to.VEMSA} deriving energy-interval estimates from the results
of the fixed-energy analysis in the \emph{continuous} $N$-particle alloy model (the situation with the discrete
$N$-particle models is simpler; cf. \cite{CS14}). In turn, Theorem \ref{thm:FEMSA.to.VEMSA}
allows us to avoid a more tedious energy-interval MPMSA procedure.

\subsection{Bounds for the flat tiling alloy model}

We start with the one-volume EVC bound, which is quite similar in form to the celebrated Wegner estimate
\cite{W81}. The flat tiling alloy model is a particular case of a much more general one, studied by
Klein and Nguyen \cite{KN13b}, so we can simply quote their result; the nature of the interaction potential
is irrelevant here. The condition \SRCM also leads to a Wegner-type estimate, but with a non-optimal
volume dependence.

\btm[Cf. \cite{KN13b}*{Theorem 2.2}]\label{thm:W1}
Let $\Sigma^{I^*}_{\Bx,L}$ be the spectrum $\Sigma(\BH_{\bcube_L(\Bx)})\cap I^*$, $I^* = [0,E^*]$. Then
\be\label{eq:thm.W1}
\pr{ \dist\big[ \Sigma^{I^*}_{\Bx,L}  , E\big] \le s } \le C_1(N,E^*, p_V) L^{Nd} \, s.
\ee
\etm

The estimate \eqref{eq:thm.W1} suffices for the fixed-energy analysis, but the derivation of dynamical (cf. \cite{C14b})
and spectral localization
requires an EVC bound for pairs of local Hamiltonians (two-volume bound). In a manner of speaking, this is
an "eigenvalue comparison" -- rather than "eigenvalue concentration" -- estimate. In the usual $1$-particle models
satisfying the condition called Independence At Distance (IAD), the "EV comparison" estimate for the pairs of spectra
in two distant finite domains follows easily from the Wegner-type estimates, but the situation is more complex with the interactive models,
particularly in a continuous configuration space.

\btm[Cf. {\cite[Theorem 4]{C14b}}]\label{thm:W2}
Under the assumptions \Vone and \Uone, for any fixed $N,d$, $E^*$ and the PDF $F_V$
of the random scatterers,
there exist some $C, A\in(0,+\infty)$ such that for any pair of
$4NL$-distant cubes $\bcube_{L}(\Bx)$, $\bcube_{L}(\By)$ the following bound holds:
\be\label{eq:thm.W2}
\forall\, s\in(0,1] \quad
\pr{ \dist\big[ \Sigma^{I^*}_{\Bx,L}, \Sigma^{I^*}_{\By,L} \big] \le s } \le C L^{A} \, s.
\ee
\etm

\section{Fixed-energy analysis of the flat tiling alloy model} \label{sec:FEMPMSA}

\subsection{Decay of the GFs in cubes}
\label{ssec:dominated.decay}

\begin{center}
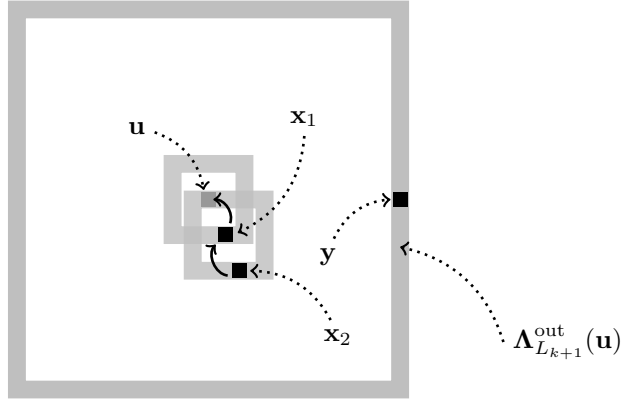
\begin{figure}
\begin{tikzpicture}
\begin{scope}[scale=0.630]
\clip (-5.0, -4.5) rectangle ++(15.0, 9.5);

\draw[color=black!50!white!50, line width = 6.7] (-4.0, -4.0) rectangle ++(8.0, 8.0);
\draw[color=black!50!white!50, line width = 6.7, opacity=0.8] (-0.75, -0.75) rectangle ++(1.5, 1.5);
\fill[color=black] (-0.15, -0.15) rectangle ++(0.3, 0.3);

\fill[color=black] ( 3.85, -0.15) rectangle ++(0.3, 0.3);

\fill[color=black] ( 0.20, -0.89) rectangle ++(0.3, 0.3);
\draw[color=black!50!white!50, line width = 6.7, opacity=0.8] (-0.33, -1.5) rectangle ++(1.5, 1.5);
\fill[color=black] ( 0.50, -1.64) rectangle ++(0.3, 0.3);

\draw[->, line width = 1.0, color=black, bend right=55] (0.45, -0.50) to (0.10, -0.00);
\draw[->, line width = 1.0, color=black, bend left=63] (0.40, -1.60) to (0.15, -0.95);

\node at (-1.5, 1.5) (Bu) {$\Bu$};
\draw[->,bend left=30, line width = 1.0, dotted] (Bu) to (-0.10, 0.2);

\node at (2.0, 1.7) (Bx1) {$\Bx_1$};
\draw[->,bend  left=30, line width = 1.0, dotted] (Bx1) to (0.61, -0.7);

\node at (2.7, -2.9) (Bx1) {$\Bx_2$};
\draw[->,bend  right=30, line width = 1.0, dotted] (Bx1) to (0.87, -1.5);

\node at (2.5, -1.2) (Bx1) {$\By$};
\draw[->,bend  left=30, line width = 1.0, dotted] (Bx1) to (3.8, -0.0);

\node at (7.5, -3.0) (BLam) {$\bcube^{\out}_{L_{k+1}}(\Bu)$};
\draw[->,bend  right=30, line width = 1.0, dotted] (BLam.west) to (4.0, -1.0);

\end{scope}
\end{tikzpicture}
\caption{\emph{Boundary belts and the GRI}.
An example of a  two-fold application of the GRI, starting with the cube $\bcube_{L_k}(\Bu)$, inside
the larger cube $\bcube_{L_{k+1}}(\Bu)$. The goal is to assess
$\|\chi_\By \BG_{\bcube_{L_k}(\Bu)} \chi_\Bu\|$. The boundary belt $\bcube^{\out}_{L_k}(\Bu)$
is covered by $|\pt^- \bball_{L_k}(\Bu)|$ cells, including $\Bx_1$. The second step
starts at the cell $\BC(\Bx_1)$ replacing $\BC(\Bu)$ used at the first step, and arrives at
the cells $\BC(\Bx_2)$, with $\Bx_2\in\pt^- \bball_{L_k}(\Bx_1)$, covering $\bcube^{\out}_{L_k}(\Bx_1)$.
The short arrows represent the factors provided by the norms $\| \cdot\|^\curlywedge$, accumulated along a descent
from the most distant cubes of radius $L_k$ down to the central cell $\BC(\Bu)$. The geometrical procedure is analogous to
the one used in the discrete models (cf., e.g., \cite[Appendix A]{CS14}).
}
\end{figure}
\end{center}

The main technical tool used here is the Geometric Resolvent Inequality, well-known in the single-particle theory and
applicable to the multi-particle Anderson Hamiltonians as well, for the structure of the potential is irrelevant for this
general analytic result. It has been used in a number of works based on the Multi-Scale Analysis of Anderson-type
operators in continuous configuration spaces; a specific implementation of this well-known technique used in the present paper
is slightly different from the standard
one\footnote{Cf., e.g., the book \cite{St01} covering the $1$-particle models,
or the paper \cite{KN13b} treating the interactive Anderson models in $\DR^d$. The geometrical constructions 
used in \cite{KN13b} are required for the bootstrap (MP)MSA deriving strong localization bounds from very weak
initial hypotheses.}. Specifically, when performing a scaling
step $L_k \rightsquigarrow L_{k+1}$, we cover the cubes of radius $L_{k+1}$ by the unit cells rather than by the central
sub-cubes $\bcube_{L_k/3}(\Bx)$ of the cubes $\bcube_{L_k}(\Bx)$. This makes the scaling scheme closer to the one
used in the framework of the discrete models and allows us to re-use, in the proof of the main analytic statement
(Lemma \ref{lem:good.NR.is.NS}) an argument
presented in our earlier works on discrete models (cf. \cites{CS13,CS14}).

Introduce a notation that will be often used below:
\be
\dnorm{ \BG_{\bcube_L(\Bu)} } := \big\| \one_{\bcubeout_L(\Bu)} \BG_{\bcube_L(\Bu)} \chi_\Bu \big\| ,
\ee
(here $\curlywedge$ symbolizes the decay from the center to the boundary of a cube),

\bpr[Cf. \cite{St01}*{Lemma 2.5.4}]
Let be given two cubes $\bcube = \bcube_\ell(\Bu)$, $\bcube' = \bcube_L(\Bv)$, such that
\be
\bcube_\ell(\Bu) \subset \bcube_{L - 3}(\Bv) \subset  \bcube_L(\Bv) .
\ee
There exists a real number $\Cgeom$
such that
for any measurable sets
$\ibA\subset \bcube^{int}_{L/3}$ and $\ibB \subset \bcube' \setminus\bcube$,
\be\label{eq:GRI.1}
\| \one_{\ibB} \BG_{\bcube}(E) \one_{\ibA} \|
\le \Cgeom
   \| \one_{\ibB} \BG_{\bcube'}(E) \one_{\bcubeout} \| \cdot \| \one_{\bcubeout} \BG_{\bcube}(E) \one_{\ibA} \|
\ee
\epr

\bco
Consider the embedded cubes $\bcube=\bcube_{L_k}(\Bx) \subset \tbcube=\bcubkoneu$ with
$|\Bx - \Bu| \le L_{k+1} - L_k - 3$.
For any cell $\BC(\By)\subset\bcube_{L_k}^{(out)}$, one has
\be
\| \chi_\By \BG_{\tbcube} \chi_\Bu \| \le \Cgeom \, \dnorm{\BG_{\bcube}} \; \| \chi_\By \BG_{\tbcube} \one_{\bcube^{\out}} \|,
\ee
and consequently (cf. \eqref{eq:def.pt.bball}), denoting $\bball = \bcube \cap \bcZ$, one has
\be\label{eq:bound.GRI.GF}
\bal
\| \chi_\By \BG_{\tbcube} \chi_\Bu \| &\le \sum_{\Bz\in \pt \bball} \Cgeom \, \dnorm{\BG_{\bcube}} \; \| \chi_\By \BG_{\tbcube} \chi_\Bz \|
\\
& \le
 \Cgeom  \, (3L)^{Nd} \dnorm{\BG_{\bcube}} \; \max_{\Bz\in \pt \bball} \| \chi_\By \BG_{\tbcube} \chi_{\Bz} \|
\eal
\ee
\eco

In \eqref{eq:bound.GRI.GF} and several formulae used below,
$(3L)^{Nd}$ is a crude upper bound on the cardinality $| \pt^- \bball_L(\Bx)|$.

As usual, we call any polynomially bounded solution to the equation
$\BH\BPsi = E\BPsi$ a generalized eigenfunction with (generalized) eigenvalue $E$.

\bpr[Cf. \cite{St01}*{Lemma 3.3.2}]\label{prop:EDI}
For spectrally a.e. $E\in\Sigma(\BH)$ there exists a generalized eigenfunction $\BPsi$, with eigenvalue $E$.
Furthermore,
\be\label{eq:EDI}
\| \chi_{\Bx} \BPsi \| \le \dnorm{ \BG_{\bcube_L(\Bx)}(E) }  \| \one_{\bcube^{\out}_L(\Bx)} \BPsi \| .
\ee
\epr

It follows from \eqref{eq:EDI} that
\be
\| \chi_\Bx \BPsi \| \le C L^{Nd} \dnorm{ \BG_{\bcube_L}(E) } \max_{ \By\in \pt^- \bball_L(\Bx)} \| \chi_\By \BPsi \| .
\ee

\bde
Let be given real numbers $m>0$, $E$ and integers $k>0$, $N\ge 1$.
A cube $\bcubeN_{L_k}(\Bu)$, as well as its lattice counterpart $\bballN_{L_k}(\Bu)=\bcubeN_{L_k}(\Bu)\cap\bcZN$, is called
$(E,m)$-non-singular (\EmNS) if
\be\label{eq:def.NS}
 \Cgeom\, (3L_k)^{Nd} \, \dnorm{ \BG_{\bcube_{L_k}\Bx}(E)} \le \eu^{ - m L_k }.
\ee
$\bcubeN_{L_{k+1}}(\Bu)$, as well $\bballN_{L_{k+1}}(\Bu)$, is called
$(E,\beta)$-NR if
\be\label{eq:def.NR}
 \dist\big( \Sigma_{\Bu,L}, \, E \big) \ge \eu^{-L^\beta}
\ee
$\bcubeN_{L_{k+1}}(\Bu)$ and $\bballN_{L_{k+1}}(\Bu)$ are called \CbetaNR if for all $L_k \le \ell \le L_{k+1} - L_k$.
\be\label{eq:def.CNR}
 \dist\big( \Sigma_{\Bu,L}, \, E \big) \ge \eu^{-L_{k+1}^\beta} .
\ee
\ede

\subsection{Induction hypothesis}
The goal of the scale induction is to prove recursively the following property:

\SS{N,k}:
Given integers $N^*\ge 3$, $L_0\ge 1$, $\alpha \ge 2$, the integer sequence
$\{L_j := (L_0)^{\alpha^j},\; j\ge 0\}$,
the real numbers $m^*>0$, $P^*>0$ and the  sequences
$$
m_n := m^*(1 + 3L_0^{-\delta+\beta})^{N^*-n}, \;\; P(n,k) := 2^k P^*(2\alpha)^{N^*-n}, \;
1 \le n \le N^*,
$$
the following property is fulfilled for all $1 \le n \le N$:
\be
\pr{ \bcube^{(n)}_{L_k}(\Bx) \text{ is \EmS } } \le L_k^{ - P(n,k) }.
\ee

It is not difficult to see that the decay of the RHS is equivalent to $\eu^{-c' \ln^{1+c} L_k}$
with $c>0$; it is faster
than any power law $L \mapsto L^{-P}$.

We summarize in the table below the assumptions on the key parameters.

\renewcommand{\arraystretch}{2.0}
\be\label{eq:table2}\hbox{\begin{tabular}{|c|c|}
  \hline
  $\tau > \max\left( \zeta^{-1}, 1 \right) $ &
      $\diy  \DN\ni \alpha  > 2 \tau  $
  \\
  \hline
\rule{0pt}{4ex}
  $\diy 0<\beta < \min\left( \frac{1}{4},\, \zeta, \, \frac{7}{8\alpha} \right) $  \rule[-3ex]{0pt}{0pt}     &  $K+1 > 4\alpha$
\\
  \hline
  $m_N = m^*\,\big(1 + 3L_0^{-1+\beta}\big)^{N^*-N} $ & $m^* \ge L_0^{-1/2}$
\\
  \hline
 $P(N,k) = 2^kP^*\, (4\alpha)^{N^*-N}$ &    $  P^* > 4 N^* d\alpha $
\\
  \hline
\end{tabular}}
\ee

Observe that
\be\label{eq:implic.for.PN}
\forall\, N=1, \ldots, N^* \quad P(N) \ge P^* > \max\big( 4Nd,  2Nd\alpha \big).
\ee

\subsection{Initial length scale (ILS) estimate}
The assumption of non-negativity of the interaction greatly simplifies the EVC analysis in the continuous multi-particle models
near the bottom of the spectrum (it is not required in the strong disorder regime, $|g|\gg 1$).
The key observation here is that any non-negative interaction can only move the EVs up, thus resulting automatically in stronger
ILS estimates (in any interval of the form $(-\infty, E^*]$)
for the interactive model at hand than with the interaction switched off.

\bpr[Cf. \cite{CBS11}*{Lemma 3.1}]
Under the assumption \Vone, there exists an interval $I^* = [0, E^*]$, with $E^*>0$, an integer $L_0$
and a real number $m^*> L_0^{-1/2}$ such that
\SS{N,0} holds true for all $1\le N \le N^*$.
\epr

\bre\label{rem:ILS}
The above ILS estimate does not require the amplitude of the random potential to be large, for
it is based on the Lifshitz tails argument, i.e., a large deviations estimate.
For the random potential of the form $gV(x;\om)$ with $g\gg 1$, one can prove \SS{N,0} with the help of a much simpler
argument, in essence going back to Ref.~\cite{DK89}, in an energy interval of length growing as $g\to +\infty$,
and with the "mass" $m_N = m_N(g)\to +\infty$.
\ere

\subsection{Analytic scaling step}

We assume that the main parameters satisfy the conditions listed in the table \eqref{eq:table2}, without repeating it every time again.
In particular, this concerns the exponent $\tau$ in the following

\bde
A cube $\bcubeN_{L_k}(\Bu)$ is called weakly interactive (WI) if
$$
\diam \Pi \Bu \equiv \max_{i\ne j} |u_i - u_j | \ge 3NL_k^\tau,
$$
and strongly interactive (SI), otherwise.
\ede

\bde
A cube
$\bcubeN_{L_{k+1}}(\Bx)$ is called \Ebad if it contains either a weakly interactive \EmS
cube of radius $L_k$ or a collection of $\ge K+1$ pairwise $9NL^\tau_k$-distant,
\EmS, strongly interactive cubes of radius $L_k$. Otherwise, it is called \Egood.
\ede

The next (deterministic) statement is a standard result of the Multi-Scale Analysis, essentially gong back to the work
\cite{DK89} and later adapted to the continuous Anderson models. The nature of the potential, in particular, the presence of interaction,
is irrelevant in the proof. The reduction to the functions $\Bx\mapsto \|\chi_\Bx \BG_{\BLam} \chi_\By\|$ with $\Bx,\By\in\bcZ^N$,
described in Sect.~\ref{ssec:dominated.decay},
allows us to apply a variant of the method presented in \cite{CS13}, but the claim can be also proved with the help of some
other known techniques; cf., e.g., Ref.~\cite{KN13b} or the book \cite{St01}.

\ble\label{lem:good.NR.is.NS}
Suppose that a cube $\bcubeLkone^{N}(\Bu)$ is \Egood and {\rm\ENR}.
If $L_0$ is large enough, then $\bcubeLkone^{N}(\Bu)$ is {\rm\EmNS}.
\ele

\proof
Set $\BLam = \bcubeLkone^{N}(\Bu)$, $\bball =\bball^{N}_{L_{k+1}-1}(\Bu)$ and fix $\By\in\pt^{-}\bball$.
Consider the function $f_\By: \bball\to\DR_+$ defined by
$$
f_\By:\Bz \mapsto\big\| \chi_\Bz \BG^{(N)}_{\BLam}(E) \chi_\By \big\|.
$$
By assumption, there is a
collection of balls $\bball(\Bu_j, L_k^\tau)\subset\BLam$, $1 \le j \le K'$, with $0\le K'\le K$, such that
any ball $\bball (\Bv,L_k )$ with $\Bv\in\bball \setminus \cup_{j=1}^{K'} \bball (\Bu_j, 9NL_k^\tau)$ is \EmNS.
Fix such a collection.
Denote by $\bcL_r(\Bu) =\{\Bz\in\bcZ :\; | \Bz - \Bu|=r\}$, $r\ge 0$ and set:
$$
\Xi :=\big\{\Bx\in\bball_{L_{k+1}-L_k-1}(\Bu):\;
\bcL_{\rd(\Bu,\Bx)}(\Bu) \cap \cup_{j=1}^{K'} \bball_{ 9NL_k^\tau}(\Bu_j) \ne \varnothing \big\}.
$$
Then any ball $\bball (\Bv,L_k )\subset\bball$
with $\Bv \in\bball\setminus \Xi$ is $(E,m_N)$-NS,
and $\Xi$ is covered by a family of $\le K$
annuli with center $\Bu$ and total width $\le K (2\cdot 9NL_k^\tau + 1) \le 19NK L_k^\tau$,
with $L_0$ large enough.
By Lemma A.2  from \cite{C14b},
$f_\By$ is $(L_k,q,\Xi)$-dominated
in  $\bball$,  in the sense of \cite[Definition A.1]{C14b},
with
$$
\bal
-\ln q &=  m_N (1 + L_k^{-1/8})L_k - L_{k+1}^{\beta} - \ln( C_\cZ^N L_{k+1}^{Nd})
\\
& \ge m_N L_k + \big(m_NL_k^{7/8} - 2 L_k^{\alpha\beta}  \big)
\ge L_k m_N \big(1 + {\textstyle\half} L_k^{-1/8} \big),
\eal
$$
where the last inequality follows from the assumptions $\alpha \beta < 7/8$ and $m_N\ge m^*\ge 1$
listed in \eqref{eq:table2}.
By  virtue of \cite[Lemma A.1]{C14b},
$$
f_\By(\Bu) \le q^{\frac{(L_{k+1}-1) - 19NK L_{k}^\tau - 1 }{L_k} } \rM(f_\By, \bball)
\le q^{\frac{ L_{k+1} - 20NK L_{k}^\tau }{L_k} } \rM(f_\By, \bball).
$$
One can see that, with $\alpha > 2\tau$, $\beta < 1/4$, $m_N\ge L_0^{-1/2} \ge L_k^{-1/2}$, and $L_0$ is large enough,
$$
\bal
\allowdisplaybreaks
-\ln f_\By(\Bu)
&  \ge  -\ln \left\{
 \left( \eu^{-m_N(1 + \half L_k^{-1/8})L_k} \right)^{\frac{L_{k+1}- 20NK L_k^\tau}{L_k}}
   \eu^{ L_{k+1}^\beta}\right\}
\\
& = m_N \left\{  \left(1+ \shalf L_{k}^{-1/8} \right)L_k \;\cdot\;
\frac{  L_{k+1} \big(1 - 20NK L_{k+1}^{-1 + \frac{\tau}{\alpha}} \big)  }{  L_{k} } - \frac{L_{k+1}^{\beta}}{m_N} \right\}
\\
& \ge m_N L_{k+1} \left\{ \left(1+ \squart L_{k}^{-1/8} \right)
   \left( 1 -  L_{k+1}^{-1/2 } \right) - L_{k+1}^{-\half - \quart}   \right\}
\\
& \ge m_N L_{k+1} \left\{ \left(1+ \squart L_{k}^{-1/8} \right)
   \left( 1 -  L_{k+1}^{-1/2 } \right) - L_{k+1}^{-3/4}    \right\}
\\
& \ge L_{k+1} m_N\left(1+ 2 L_{k+1}^{-1/8} \right)
 \ge  \gamma(m_N, L_{k+1})L_{k+1} + \ln \big( C_{\cZ,N}L_{k+1}^{Nd} \big) .
\eal
$$
Therefore, $\bcube_{L_{k+1}}(\Bx)$ is \EmNS, with the same value of $m^*$ figuring in $m_N = m^*(1 + 3L_0^{-1+\beta})^{N^* - N}$, i.e.,
$m^*\ge L_0^{-1/2}$.
This completes the proof.
\qedhere

\subsection{Probabilistic scaling step}

\subsubsection{Weakly interactive cubes }

\ble\label{lem:WI.decomp}
For any weakly interactive cube $\bcubeN_{L_k}(\Bu)$ there is a factorization
$\bcubeN_{L_k}(\Bu) = \bcube^{n'}_{L_k}(\Bu') \times \bcube^{n''}_{L_k}(\Bu'')$
such that
\be
\dist\big( \Pi \bcube^{n'}_{L_k}(\Bu'), \, \Pi \bcube^{n''}_{L_k}(\Bu'') \big) > L^\tau.
\ee
\ele

\proof
Assuming $\diam( \Pi \Bu) > 3 N L^\tau$, let us show that the projection $\Pi \bcube(\Bu,3L/2)$ is a disconnected subset of $\cZ$.
Assume otherwise; then for any partition $\cJ \sqcup \cJ^\rc = \{1, \ldots, N\}$,
we have $\rd(\Pi_\cJ \Bu, \Pi_{\cJ^\rc}\Bu)$   $\le 2 \cdot \frac{3L^\tau}{2}=3L^\tau$. Then
it is readily seen that
$\diam\; \Pi\Bu \le (N-1) \cdot 3L^\tau < 3NL^\tau$, contrary to our hypothesis.
Therefore, one has
$\rd\left( \Pi_\cJ \bcube_{3L^\tau/2}(\Bu),  \Pi_{\cJ^\rc} \bcube_{3L^\tau/2}(\Bu)\right) > 0$,
for some partition $(\cJ,\cJ^\rc)$,
hence
$$
\rd\left( \Pi_\cJ \bcube_{3L^\tau/2}(\Bu),  \Pi_{\cJ^\rc} \bcube_{3L^\tau/2}(\Bu)\right)>  \half L^\tau + \half L^\tau = L^\tau ,
$$
as asserted.
\qedhere

We will assume that one such factorization is associated with each WI cube (even if it is not unique), and call
it the canonical one. For the Hamiltonian in a WI cube we have the following algebraic representation:
with $\bcube' = \bcube^{n'}_{L_k}(\Bu')$, $\bcube''=\bcube^{n''}_{L_k}(\Bu'')$,
\be
\bal
\BH &= \BHni + \BU_{\bcube', \bcube''}
\\
& = \BH_{\bcube'} \otimes \one^{(n'')} + \one^{(n')} \otimes \BH_{\bcube''} + \BU_{\bcube', \bcube''}
\eal
\ee
where, due to the assumption \Uone,
\be\label{eq:Uone.norm.Ubcube.bcube}
\| \BU_{\bcube', \bcube''} \| \le C  \eu^{-L_k^{\tau\zeta} }, \;\; \text{ with $\tau\zeta>1$ by \eqref{eq:table2}. }
\ee

\ble\label{lem:prob.WI.S}
Assume the property {\rm\SS{N-1,k}}. If $L_0$ is large enough, then for any WI cube $\bcubeN_{L_k}(\Bu)$
\be\label{eq:lem.prob.WI.S.1}
\pr{ \bcubeN_{L_k}(\Bu) \text{ is \EmS } } \le L_k^{ - \frac{3}{2}  \,P(N,k+1) }
\ee
and therefore,
\be\label{eq:lem.prob.WI.S.2}
\bal
\pr{ \bcubeN_{L_{k+1} }(\Bu) \text{ contains  a WI \EmS ball of radius $L_k$}}
\le \frac{1}{4} L_k^{ - P(N,k+1) } .
\eal
\ee
\ele

See the proof in Appendix \ref{app:proof.WI.S}.

\subsubsection{Strongly interactive cubes}

\ble[Cf. {\cite[Lemma 3.]{C14b}}]\label{lem:8NL.dist}
If two {\rm SI} cubes $\bcubeN_L(\Bx)$, $\bcubeN_L(\By)$ are $9NL^\tau$-distant
and $L > 2\rr_1$, then
\be
\Pi \bcubeN_{L+\rr_1}(\Bx) \cap \Pi \bcubeN_{L+\rr_1}(\By) = \vempty
\ee
and, consequently, the random operators $\BH_{\bcubeN_L(\Bx)}$ and $\BH_{\bcubeN_L(\By)}$
are independent.
\ele

\subsubsection{The scale induction}

\btm
Suppose that \SS{N,0} holds true, and for all $k\ge 0$, one has
$$
\pr{ \text{ $\bcube_{L_{k+1}}(\Bu)$ is \ER } } \le  \quart L_{k+1}^{-P(N,k)} .
$$
If $L_0$ is large enough, then \SS{N,k} holds true for all $k\ge 0$.
\etm

\proof
It suffices to derive \SS{N,k+1} from \SS{N,k}.
By Lemma \ref{lem:good.NR.is.NS}, if $\bcube_{L_{k+1}}(\Bu)$ is \EmS, then either it is not \CbetaNR, or it is \Ebad. Let
\begin{align}
\notag
\rP_i &:= \pr{ \text{ $\bcube_{L_{k+1}}(\Bu)$ is \EmS} }, \;\; i=k, k+1,
\\
\notag
\rS_{k+1} &:= \pr{ \text{ $\bcube_{L_{k+1}}(\Bu)$ contains a WI, \EmS cube of radius $L_k$} },
\\
\label{eq:Q.k+1}
\rQ_{k+1} &:= \pr{ \text{ $\bcube_{L_{k+1}}(\Bu)$ is \ER } } \le \quart L_{k+1}^{-P(n,k+1)} ,
\end{align}
(the last inequality is assumed, but its validity actually follows from Theorem \ref{thm:W1}).
Further, an \Ebad cube $\bcube_{L_{k+1}}(\Bu)$ must contain either a WI, \EmS cube of radius
$L_k$ (with probability
$\rS_{k+1}\le \quart L_{k+1}^{-P(n,k+1)}$
by Lemma \ref{lem:prob.WI.S}), or at least $K+1$ pairwise $9NL^\tau_k$-distant cubes
$\bcube_{L_k}(\Bv_i)$, $i=1,\ldots, K+1$, which are \EmS. By Lemma \ref{lem:8NL.dist}, the random operators
$\BH_{\bcube_{L_k}(\Bv_i)}(\om)$ are independent, thus such an event occurs with probability
$$
\le C L_{k+1}^{(K+1)Nd} \rP_k^{K+1}
\le C L_k^{ - (K+1)\big[P(N) -Nd\alpha \big] }
 \le \quart L_{k+1}^{ - 2 P(N,k) } = \quart L_{k+1}^{ -P(N,k+1) } ,
$$
under the conditions $P(N) > 4Nd\alpha$, $K+1 \ge 4\alpha$
given in the table \eqref{eq:table2}.
Therefore,
$$
\bal
\rP_{k+1} &\le C L_{k+1}^{(K+1) Nd} \rP_k^{K+1} + \rS_{k+1}+  Q_{k+1}
\\
& \le \quart L_{k+1}^{-P(n,k+1)}  + \quart L_{k+1}^{-P(n,k+1)}  + \quart L_{k+1}^{-P(n,k+1)}
<  L_{k+1}^{-P(n,k+1)}  .
\eal
$$
\qedhere

\section{Exponential spectral localization }
\label{sec:reduction}

\subsection{Energy-interval MPMSA estimates}
\label{ssec:VEMPMSA.exp}

Introduce the following notation which will be used in this subsection:
$$
\BF_\Bx(E) = \BF_{\Bx,L}(E) := \max_{ \Bz \in\pt^- \bball_{L}(\Bx) }
   \left\| \chi_\By \BG_{\bcube_{L}(\Bx)}  \chi_\Bx  \right\| .
$$

\btm\label{thm:FEMSA.to.VEMSA}
Fix $L\ge 1$ and assume that the following fixed-energy bound holds true for some $a_L, q_L>0$:
$$
\forall\, E\in I^* \qquad \pr{ \BF_\Bx(E) \ge a_L} \le q_L.
$$
Assume also that
%
the EVC bound  of the form \eqref{eq:thm.W2} holds true for a pair
of cubes $N$-particle cubes $\bcube_{L}(\Bx)$, $\bcube_{L}(\By)$.
Then for any $b>0$, one has
\be\label{eq:thm.2vol.3NL.balls}
\pr{\exists\, E\in I^*:\,  \min(\BF_\Bx(E), \BF_\By(E)) \ge a_L } \le 2|I^*| b^{-1} q_L + C''' L^{4Nd} b .
\ee

In particular, under the assumption  \Vone, the bound \eqref{eq:thm.2vol.3NL.balls} holds true for any pair of $4NL_k$-distant
cubes of radius $L_k$, owing to Theorem
\ref{thm:W2}.
\etm

\begin{proof}

By the Chebychev inequality
combined with the Fubini theorem, we have
\be\label{eq:Fubini.x}
\bal
\pr{ \cS_{b,\Bx} } & \le
b^{-1}\esm{ \int_{I } \one_{\{ \BF_\Bx(E)\ge a \} } \, dE }
= b^{-1}\int_{I } \esm{\one_{\{\BF_\Bx(E)\ge a \} }  }\, dE
\\
& = b^{-1} \int_{I } \pr{\BF_\Bx(E)\ge a }  \, dE
\le b^{-1} |I | q_L.
\eal
\ee
Similarly,
\be\label{eq:Fubini.y}
\bal
\pr{ \cS_{b,\By} } & \le b^{-1} |I | q_L.
\eal
\ee
For any $\om\not\in \cB_b := \cB_{b,\Bx} \cup \cB_{b,\By}$, each of the sets
$$
\csE_\Bx(a) := \{E\in I:\, \BF_\Bx > a\}, \; \csE_\By(a) := \{E\in I:\, \BF_\By > a\},
$$
has Lebesgue measure bounded by $b$. The norm of the resolvent is a continuous function of the energy $E$,
on the complement to the spectrum, and the latter is discrete for any finite volume Hamiltonian,
thus the set $\{E\in (0,E^*): \, \BF_\Bx>a\}$ is decomposed into open sub-intervals;
the same is true for $\BF_\By$. Therefore,
$$
\begin{array}{ll}
\csE_\Bx(a) &= \cup_{i=1}^{K'} J_{\Bx,i}, \;\; \sum_{i=1}^{K'} |J_{\Bx,i}| \le b, \;\;\;\; K'\le +\infty,
\\
\csE_\By(a) &= \cup_{j=1}^{K''} J_{\By,j}, \;\; \sum_{j=1}^{K''} |J_{\By,j}| \le b, \;\;\;\; K''\le +\infty ,
\end{array}
$$
and
$$
\bal
\pr{ \exists\, E\in I^*:\, \min\big[ \BF_\Bx(E), \BF_\By(E)\big] >a} &\le
\pr{ \csE_\Bx(a) \cap \csE_\By(a) \ne \varnothing}
\\
& \le \sum_{i=1}^{K'} \sum_{j=1}^{K''} \pr{ J_{\Bx,i} \cap J_{\By,i} \ne \varnothing} .
\eal
$$
Denote $\eps_{\Bx,i} := |J_{\Bx,i}|$, $\eps_{\By,i} := |J_{\By,i}|$; then for any fixed  pair $(i,j)$, we have
$$
\bal
\pr{ J_{\Bx,i} \cap J_{\By,i} \ne \varnothing} &\le \pr{ |\xi - \mu_{ij}| \le |n_1 - n_2|^{-1} (\eps_{\Bx,i} + \eps_{\By,i}) }
\\
& \le C L^{A} \left(\eps_{\Bx,i} + \eps_{\By,i} \right),
\eal
$$
by virtue of Proposition \ref{prop:SRCM}, thus
$$
\pr{ \exists\, E\in I^*:\, \min\big[ \BF_\Bx(E), \BF_\By(E)\big] >a}
\le C L^{A} \sum_{i=1}^{K'} \sum_{j=1}^{K''} \left(\eps_{\Bx,i} + \eps_{\By,i} \right)
\le  C L^{A} \cdot 2b.
$$

\end{proof}

\subsection{Exponential decay of eigenfunctions. Proof of Theorem \ref{thm:main}}

In the next statement, we keep the same notations for the Hamiltonian and the cubes as before, but it can be easily seen
that the result applies to a much larger class of Schr\"{o}dinger operators in a Euclidean space $\DR^D$, $D\ge 1$,
with bounded measurable random potential $\DR^D \ni x\mapsto W(x;\om)$. In our case, $D = Nd$, $x$ is replaced by $\Bx$, and
$W(\Bx;\om) = \BV(\Bx;\om) + \BU(\Bx)$. The constant $a$ figuring in Lemma \ref{lem:VEMSA.implies.SL}
can be set to $4N$, owing to Theorem \ref{thm:W2}. The main argument is not new, and we present it here only for completeness.
Its structure is very close to the one employed in a number of papers
on the single-particle MSA, thanks to the bound \eqref{eq:VEMSA.bound} established for all pairs of cubes which are $aL_k$-distant in the
norm-distance, and not in the Hausdorff distance (cf. \cites{CBS11,KN13b}).

\ble\label{lem:VEMSA.implies.SL}
Consider the random Hamiltonian $\BH(\om)$ and assume that for some $a\in(0,+\infty)$ and interval $I^*\subset \DR$,
for any $k\ge 0$ and any pair of $aL_k$-distant cubes $\bcube_{L_k}(\Bx)$, $\bcube_{L_k}(\By)$,
the following probabilistic bound holds true:
\be\label{eq:VEMSA.bound}
\pr{\exists\, E\in I^*:\,  \text{ $\bcube_{L_k}(\Bx)$ and $\bcube_{L_k}(\By)$ are \EmS} } \le L_k^{-p_k}
\ee
where $\lim_k p_k = +\infty$.
Then with probability one, every nontrivial polynomially bounded solution $\BPsi$ to the equation $\BH(\om)\BPsi = E\BPsi$ with $E\in I^*$ decays exponentially
fast at infinity, with rate $\ge m^*>0$. Specifically, for some $r(\BPsi)\in(0,+\infty)$ and all $\Bx\in\bcZ^N$ with $|\Bx|\ge r(\BPsi)$, one has
\be\label{eq:lem.VEMSA.implies.SL}
\| \chi_\Bx \BPsi \| \le \eu^{ - m^*|\Bx|}.
\ee
\ele

\begin{proof}
Fix a polynomially bounded solution $\BPsi$ which is not a.e. zero, so there exists $\hBx\in\bcZN$ such that $\| \chi_{\hBx} \BPsi \|>0$.
Fix such a point $\hBx$.

Furthermore, there exists an
integer $k_0\ge 0$ such that for all $k\ge k_0$, $\bcube_{L_k}(\hBx)$ is \EmS.
Assume otherwise, then there are arbitrarily large cubes $\bcube_{L_k}(\hBx)$
such that
$$
\|  \chi_{\hBx}\Psi\| \le C_1 L_k^{C_2} \eu^{-m_N L_k} \tto{L_k\to\infty} 0,
$$
which contradicts our assumption that $\| \chi_{\hBx}\Psi \| >0$.

We fix $k_0$ and work only with $k\ge k_0$ and denote $\DA_k := \bcube_{2L_{k+2}}(\Bzero) \setminus \bcube_{aL_{k}}(\Bzero)$.
Introduce the event
$$
\cT_k(\BLam) :=
\{\exists\, E\in I^*: \,  \BLam \text{ contains two $aL_k$-distant \EmS cubes  } \}
$$
By the assumed property \eqref{eq:VEMSA.bound},
$$
\bal
\pr{ \cT_k(\bcube_{2L_{k+2}}(\Bzero) )  } & \le  |\bcube_{2L_{k+2}}(\Bzero)|^2\,  L_k^{-p_k}
\le C L_k^{ - p_k + P^*2Nd\alpha^2}
\\
& \le C' L_k^{ - p_k/2} .
\eal
$$
The last RHS is summable in $k \ge k_0$, so by the Borel--Cantelli lemma, there is a subset $\tOm\subset\Om$ with $\pr{\tOm}=1$
such that for any $\om\in\tOm$ there exists
$k_1\ge k_0$ such that for all $k\ge k_1$, the event $\cT_k(\bcube_{2L_{k+}}(\Bzero) )$ does not occur.
Since $\DA_k\subset \bcube_{2L_{k+}}(\Bzero)$, and for all $k \ge k_1\ge k_0$, the cube $\bcube_{L_{k}}(\Bzero)$
is \EmS, all cubes $\bcube_{L_k}(\By)\subset \DA$ (with $k\ge k_1$) are \EmNS.

Fix $\om\in\tOm$. Now the argument becomes deterministic.

Fix any $\Bx$ with $|\Bx|> aL_{k_1}$, and let $k = k(|\Bx|)\in\DN$ be such that $|\Bx| \in(2L_{k+1}, 2L_{k+2}]$.
Consider the cube $\bcube_{|\Bx| - aL_k}(\Bx)\subset \DA_k$. It follows from the choice of $k_1 \,(\le k)$ that
all cubes of radius $L_k$ inside $\bcube_{|\Bx|- aL_k}(\Bx)$ are \EmNS.
Therefore, by Proposition \ref{prop:EDI}, using recursively the \EmNS property of the cubes of radius $L_k$ inside
$\bcube_{|\Bx|- 4NL_k}(\Bx)$ and taking into account that $|\Bx| > L_{k+1} = L_k^\alpha$, we obtain:
$$
\bal
- \ln \| \chi_\Bx  \BPsi \| &\ge  \gamma(m_N, L_k) (|\Bx| - aL_k ) - C_3 \ln L_k
\\
& \ge m_N(1 + L_k^{-1/8}) |\Bx| \left(1 - 4N L_k^{1 - \alpha} - C L_k^{-\alpha} \ln L_k \right)
\\
& \ge m_N |\Bx| \left(1 + \half L_k^{-1/8} \right) \left(1 - 5N L_k^{-1} \right) \;\; \text{ (with $\alpha > 2$) }
\\
& \ge m_N |\Bx|  \ge m^* |\Bx|.
\eal
$$
In other words, there exists $r(\BPsi)<\infty$ such that for all $\Bx\in\bcZ^N$ with $|\Bx|\ge r(\BPsi)$,
\be\label{eq:lem.VEMSA.implies.SL.again}
\| \chi_\Bx \BPsi \| \le \eu^{- m^* |\Bx|}.
\ee
\end{proof}

\proof[Proof of Theorem \ref{thm:main}]
By Proposition \ref{prop:EDI}, for spectrally a.e. $E\in\DR$ there exists a generalized eigenfunction $\BPsi$  with (generalized) eigenvalue
$E$. By Lemma \ref{lem:VEMSA.implies.SL}, every generalized eigenfunction $\BPsi$ with eigenvalue in $I^*$ is square-summable,
hence the spectrum of $\BH(\om)$ in $I^*$ is pure point, and
there is a countable family of $L^2$-eigenfunctions $\BPsi_j(\om)$ of $\BH(\om)$ with eigenvalues in $I^*$.
Now the claim follows from \eqref{eq:lem.VEMSA.implies.SL.again}.
\qedhere

\appendix

\section{Proof of Lemma \ref{lem:prob.WI.S}}
\label{app:proof.WI.S}

\ble\label{lem:WITRONS.subexp}
Fix $\beta \in (0,1]$, $m^*\geq 1$ and $E\in\DR$.
Suppose that a {\rm WI} ball $\bballN (\Bu,L_k)$
is $(E,\beta)$-NR and
satisfies the following two conditions:
\begin{align}
\label{eq:PNS.1}
\forall\, &\lam' \in I^* \cap \Sigma\left(\BH^{(N')}_{\bball'}\right) \quad
\;\,\bball'' \text{ is } (E-\lam', m_{N'})-{\rm{NS}} ,
\\
\label{eq:PNS.2}
\forall\, &\lam'' \in I^* \cap \Sigma\left(\BH^{(N'')}_{\bball''}\right) \quad
 \bball' \text{ is } (E-\lam'', m_{N''})-{\rm{NS}}.
\end{align}

If $L_0$ is large enough
then  $\bball^{(N)}(\Bu,L_k)$ is {\rm$(E, m_{N})${\rm-NS}}.
\ele

\proof
The operator $\BH_{\bcube'}$ has compact resolvent, thus its eigenvalues $E'_a\uparrow +\infty$ as $a\to+\infty$.
Similarly, for the eigenvalues $E''_a$ of $\BH_{\bcube'}$ we have $E''_a\uparrow +\infty$ as $a\to+\infty$.
The EVs of the operators appearing in our arguments are assumed to be numbered in increasing order.
We have the following identities:
\be\label{eq:Gni.proj.psi}
\BG_{\bcube_{L_k}(\Bu)}(E) = \sum_{a} \BP'_{\BPsi'_a}  \otimes \BG_{\bcube''}(E - E'_a)
= \sum_{a} \BG_{\bcube'}(E - E''_a) \otimes \BP''_{\BPsi''_a} ,
\ee
where $\BPsi'_a$ are the eigenfunctions of $\BH^{(N')}_{\bball'}$ and $\BPsi''_a$ the eigenfunctions of $\BH^{(N'')}_{\bball''}$.

By the second resolvent identity, for any energy $E$ which is not in the spectra of $\BHni_{\bcube}$ and $\BH_{\bcube}$,
we have for their respective resolvents $\BGni_\bcube(E)$ and $\BG_\bcube(E)$
$$
\BG_{\bcube} = \BGni_{\bcube} - \BGni_{\bcube} \BU_{\bcube',\bcube''} \BG_{\bcube}
$$
thus
$$
\bal
\| \chi_\By \BG \chi_\By  \| &\le \|\chi_\By \BGni \chi_\Bx \|  + \| \chi_\By \BGni \BU \BG \chi_\Bx \|
\\
& \le \|\chi_\By \BGni \chi_\Bx \|  + \| \BU_{\bcube',\bcube''}\| \| \BGni_{\bcube}\| \| \BG_{\bcube}\|.
\eal
$$

We start with the last term in the RHS.
Since $\BLam$ is weakly interactive, we have by
inequality \eqref{eq:Uone.norm.Ubcube.bcube} (cf. also  Lemma \ref{lem:WI.decomp}), with $\tau\zeta >1$ by \eqref{eq:table2},
$$
\| \BU_{\bcube',\bcube''} \| \le C \eu^{- (3NL^\tau_k)^\zeta} < C' \eu^{ - c L_k^{\tau\zeta}} < \eu^{ - \tm L_k },
$$
where $\tm>0$ can be made arbitrarily large, provided $L_0$ is large enough. It suffices to have, e.g., $\tm > 3m_N$, which we assume below.

The assumed $(E,\beta)$-NR property gives $\| \BG_{\bcube}\| \le \half \eu^{L_k^{\beta}}$; in terms of the spectrum
$\Sigma_\bcube$ of $\BH_{\bcube}$,
$\dist(E, \Sigma_\bcube)\ge 2\eu^{-L_k^{\beta}}$. The min-max principle implies for
the spectrum $\Sigmani_\bcube$ of $\BHni_{\bcube}$
\be\label{eq:perturbed.nonres.subexp}
\dist(E, \Sigmani_\bcube)\ge 2\eu^{-L_k^{\beta}} - \| \BU_{\bcube',\bcube''}\| \ge \eu^{-L_k^{\beta}},
\ee
so $\| \BGni_{\bcube}\| \le \eu^{L_k^{\beta}}$. Finally, with $\tm > 3m_N$,
$$
\| \BU_{\bcube',\bcube''}\| \| \BGni_{\bcube}\| \| \BG_{\bcube}\| \le  C \eu^{- \tm L_k + 2 L_k^\beta}
\le \half \eu^{- m_N L_k }.
$$

It remains to assess the GF of the non-interacting Hamiltonian.
Denote
$$
a(\eta) = \max\{a: \, E'_a \le E_* + 2\eta\},
$$
and set $\eta = m_N$.
It follows from the Weyl law that $\card \{a: \, E''_a \le E_* + 2\eta\}\le L_k^{C}$,
for some $C = C(d,N)<+\infty$.
The Combes-Thomas estimate (cf. \cites{CT73,St01}) implies that
$$
\sum_{a > a(\eta) } \dnorm{ \BP'_{\BPsi'_a} \otimes \BG_{\bcube''}(E - E'_a) }
\le \sum_{j=1}^{+\infty} L_k^{C} \eu^{ - (2\eta + j)L_k } \le \frac{1}{2} \eu^{ - \eta L_k}
\le \frac{1}{2} \eu^{ - 2m_N L_k } .
$$
By assumption, for all $a\le a(\eta)$,
$$
\dnorm{ \BG_{\bcube''}(E - E'_a) }  \le \eu^{ - m_{N-1} L_k } \le \eu^{ - 2m_{N} L_k }.
$$
We conclude that
\be\label{eq:WI.NS.1}
\bal
\sum_{a}
\big\| \chi_{\By}  \BP'_{\BPsi'_a} \otimes \BG_{\bcube''}(E - E'_a) \chi_{\By} \big\|
  & \le \left( \sum_{a \le a(\eta)} + \sum_{a > a(\eta)} \right) \BP'_{\BPsi'_a}  \otimes \BG_{\bcube''}(E - E'_a)
\\
& \le L_k^{C'} \, \eu^{ - 2m_{N} L_k } + \frac{1}{2} \eu^{ - 2m_N L_k }
\le \eu^{ - 2 m_N L_k }.
\eal
\ee
Similarly,
\be\label{eq:WI.NS.2}
\bal
\sum_{a}
\big\| \chi_{\By}  \BG_{\bcube'}(E - E''_a)\otimes \BP''_{\BPsi''_a} \chi_{\By} \big\|
  & \le \eu^{ - 2 m_N L_k }.
\eal
\ee
Taking the sum over all $\By\in \pt^{-} \bball_{L_k}(\Bu)$, falling into one of the two categories
\eqref{eq:WI.NS.1}--\eqref{eq:WI.NS.2}, we obtain for $L_0$ large enough
$$
\dnorm{ \BG_{\bcube_{L_k}(\Bu)}(E) } \le \Const L_k^{Nd} \eu^{ -2 m_N L_k }
\le \eu^{ -m_N L_k },
$$
which proves the assertion of the lemma.
\qedhere

\proof[Proof of Lemma \ref{lem:prob.WI.S}]

Denote by $\cS$ the event in the LHS of \eqref{eq:lem.prob.WI.S.1}. Let $\bcube=\bcube^{(N)}(\Bu,L_k)$ and
consider the canonical factorization $\bcube=\bcube'\times\bcube''$.
We have
\be\label{eq:proof.lem.WI.T.a}
\bal
\pr{\cS } & < \pr{\text{ $\bcube$  is not  \ENR }}
\\
&
+ \pr{\text{ $\bcube$  is  \ENR and \EmS }}.
\eal
\ee
By Theorem \ref{thm:W1}, the first term in the RHS of \eqref{eq:proof.lem.WI.T.a} is bounded by
$\eu^{-L_{k+1}^{\beta}}$, so we focus on the second summand.

Let $\Sigma' =\Sigma\big(\BH^{(N')}_{\bball'}\big)\cap I^*$,
$\Sigma'' =\Sigma\big(\BH^{(N'')}_{\bball''}\big)\cap I^*$,
and consider the events
$$
\bal
\cS' &= \{\om:\, \exists\, \lam' \in \Sigma', \;
\bball'' \text{ is } (E-\lam', m_{N'})-{\rm{NS}} \} ,
\\
\cS'' &= \{\om:\,  \exists\, \lam'' \in\Sigma'', \;
 \bball' \text{ is } (E-\lam'', m_{N''})-{\rm{NS}} \}.
\eal
$$

Notice that, although the spectra $\Sigma', \Sigma''\subset I^*$ are random, their cardinalities are bounded
by those for the respective Laplacians, with the potential energy $\BV + \BU$ switched off, owing to the positivity of the latter.
These cardinalities are polynomially bounded in $L_k$, by the Weyl law.

Since $\bcube$ is WI, we have that
$\Pi \bcube' \cap \Pi \bcube'' = \varnothing$, $\BH_{\bcube''}(\om)$ is independent of
the sigma-algebra $\fF'$ generated by the random scatterers affecting $\bcube'$, while
$\BH_{\bcube'}(\om)$ is $\fF'$-measu\-rable, and so are all the EVs $\lam'\in\Sigma'$.

Further, by non-negativity of $\BH'$, if $E \le E^*$, then $E - \lam' \le E^*$ for all $\lam' \in \Sigma'$.

Replacing the quantity $E-\lam'$, rendered nonrandom by conditioning on $\fF'$, with a new nonrandom
parameter $E'\le E^*$, we have  by induction in $1 \le n \le N-1$
\be\label{eq:prob.cS.1}
\bal
\pr{\cS'} &= \esm{ \pr{\csS' \,|\, \fF''} } \le \sup_{ E' \le E^*} \pr{ \bcube'' \text{ is $(E',m)$-S}}
\\
& \le C |\bcube''|\, L_k^{ - P(N-1,k) } \le C' \, L_k^{ - 4\alpha P(N,k) + Nd}
\le \third  L_{k+1}^{ - 4P(N,k) + Nd\alpha^{-1}}  .
\eal
\ee
Using the definition of $P(N,k)$ in \eqref{eq:table2}, we have
$$
4 P(N,k) = 4 \cdot 2^k P^* (2\alpha)^{N^*-N} = 2 P(N,k+1),
$$
so
$$
4P(N,k) - Nd \alpha^{-1} = 2 P(N,k+1) - \half Nd  > \frac{3}{2} P(N,k+1) ,
$$
since $P(N,k+1)\ge P^* > 4Nd$ (cf. \eqref{eq:table2}). Thus
\be\label{eq:prob.cS.1.again}
\bal
\pr{\cS'}  \le \third L_{k+1}^{ - \frac{3}{2} P(N,k+1)}
\eal
\ee
and, similarly,
\be\label{eq:prob.cS.2}
\bal
\pr{\cS''}  \le \third L_k^{ - \frac{3}{2} P(N,k)}  .
\eal
\ee
Collecting \eqref{eq:proof.lem.WI.T.a}--\eqref{eq:prob.cS.2},
the assertion \eqref{eq:lem.prob.WI.S.1} follows.

For the second assertion \eqref{eq:lem.prob.WI.S.2}, it suffices to apply a polynomial bound $C L_{k+1}^{Nd}$ on the number of
cubes of size $L_k$ with centers on the lattice $\bcZ^N$ in a cube of radius $L_{k+1}$.
\qedhere


\end{document}